\documentclass[11pt]{article}
\newcommand\be{\begin{equation}}
\newcommand\ee{\end{equation}}
\newcommand\fa{\begin{eqnarray}}
\newcommand\ffa{\end{eqnarray}}
\usepackage[dvips]{graphicx}
\def\Comp{{\mathchoice
{\setbox0=\hbox{$\displaystyle\rm C$}\hbox{\hbox to0pt
{\kern0.4\wd0\vrule height0.9\ht0\hss}\box0}}
{\setbox0=\hbox{$\textstyle\rm C$}\hbox{\hbox to0pt
{\kern0.4\wd0\vrule height0.9\ht0\hss}\box0}}
{\setbox0=\hbox{$\scriptstyle\rm C$}\hbox{\hbox to0pt
{\kern0.4\wd0\vrule height0.9\ht0\hss}\box0}}
{\setbox0=\hbox{$\scriptscriptstyle\rm C$}\hbox{\hbox to0pt
{\kern0.4\wd0\vrule height0.9\ht0\hss}\box0}}}}
\def\C{\Comp}

\def\cc{\circ}

\def\Si{\Sigma}
\def\G{\Gamma}

\def\b{\beta}

\def\e{\epsilon}

\def\a{\alpha}
\def\d{\delta}

\def\t{\tau}
\begin{document}

\title{Quantization of static space-times}
\author{Yongge Ma\thanks{E-mail: ma@gravity.phys.psu.edu} \thanks{New address:
Dept of Physics, Beijing Normal University, Beijing 100875, China.} \\
\centerline{\textit{Center for Gravitational Physics and Geometry}}\\
\centerline{\textit{The Pennsylvania State University}}\\
\centerline{\textit{University Park, PA 16802, USA}.}}

\maketitle

\begin{abstract}
\baselineskip=20pt

A 4-dimensional Lorentzian static space-time is equivalent to
3-dimensional Euclidean gravity coupled to a massless Klein-Gordon
field. By quantizing canonically the coupling model in the
framework of loop quantum gravity, we obtain a quantum theory
which actually describes quantized static space-times. The
Kinematical Hilbert space is the product of the Hilbert space of
gravity with that of imaginary scalar fields. It turns out that
the Hamiltonian constraint of the 2+1 model corresponds to a
densely defined operator in the underlying Hilbert space, and
hence it is finite without renormalization. As a new point of
view, this quantized model might shed some light on a few physical
problems concerning quantum gravity.

\end{abstract}

\newpage

\section{Introduction}
\label{sec:1} \baselineskip=20pt

Since the Ashtekar variables was proposed in 1986 \cite{As86},
considerable progress has been made in non-perturbative canonical
quantum gravity, namely, loop quantum gravity or quantum geometry
\cite{Ro98,As98}. The kinematics of this theory has been
rigorously defined \cite{AL95,Ba96}. Certain geometric operators
corresponding to the measurement of length, area, volume, and the
integrated norm of any smooth one forms are shown to have discrete
spectra \cite{Th96a,RS95,AL97,DR96,AL98,ML00a}. The classical
limit of the quantum theory is currently under investigations
\cite{ARS92,ML00b,CR01,Bo01,Th01}. Moreover, the fundamental
discreteness in loop quantum gravity is crucially used to make
much new progress such as: the derivation of black hole entropy
from loop quantum gravity \cite{Ro96,AB98,AB00}, the resolution of
the big-bang singularity in loop quantum cosmology \cite{Boj01},
and the proof of the finiteness in the path integral, namely
spin-foam, approach to quantum gravity \cite{Pe01,CP01}.

Despite these achievements, some important elements in this
approach are yet to be understood. Although the Barbero-Immirzi
parameter \cite{Im97} could be crucially selected in the
calculations of black hole entropy in order to match the
semi-classical result of Bekenstein and Hawking, some aesthetic
criticisms are raised against the real connection formulation of
Lorentz gravity \cite{Sa00}. Despite the systematic efforts toward
constructing Hamiltonian operators in the underlying Hilbert space
\cite{Th96,BG99}, the dynamics of the quantum theory has not been
fully understood, especially if one wants to consider
asymptotically non-flat cases. The primary goal of this paper is
to test the loop quantization techniques by considering a kind of
simplified model, which are static space-times. Since static
solutions to the vacuum Einstein equation are equivalent to
3-dimensional Euclidean gravity coupled to massless Klein-Gordon
fields, the "Immirzi ambiguity" would be avoided in the model. It
is known that 3+1 Lorentz gravity coupled to Higgs scalar fields
could be quantized in the framework of loop quantum gravity
\cite{Th97b}. While, since that construction of the Hamiltonian
constraint operator, which reflects the dynamics, depends
crucially on the detail structure of 3+1 dimension, it is not
obvious if a similar construction is still available in 2+1
dimension. Meanwhile, the canonical treatment of the Euclidean
3-space could provide a possibility to address the quantization of
surface terms arising from the 3+1 Lorentz gravity. Moreover, in
this framework one could expect to calculate black hole entropy by
counting the number of the physical states on an apparent horizon.

In Section 2, paying attention to the necessary background for
further expositions, we briefly review the main framework of
canonical quantization of 2+1 Euclidean gravity as well as Higgs
fields coupled to gravity, which is developed by Thiemann
\cite{Th97a,Th97b}. We show in Section 3 that 4-dimensional
Lorentz static spacetimes are equivalent to 3-dimensional
Euclidean gravity coupled to massless Klein-Gordon fields.
Static space-times are then quantized in
section 4 by canonically quantizing the model of Euclidean
3-gravity coupled to the scalar fields. The kinematical Hilbert
space of the model is the product of the Hilbert space of gravity
and that of imaginary scalar fields, ${\cal H}_E\otimes{\cal
H}_S$. The Hamiltonian constraint is successfully quantized as a
densely-defined operator. Similar to the case in 3+1 dimension, it
is completely finite without renormalization. Section 5 addresses
a few directions for future investigations. The operator
corresponding to the area of any oriented 2-manifold is
constructed under certain circumstances, and its self-adjointness
is proved. This operator is supposed to be useful for future
applications of this framework, including a possible quantization
of the surface terms in the gravitational Hamiltonian.

\section{2+1 Euclidean quantum gravity and Higgs field coupling}
\label{sec:2}

\subsection{3-dimensional Euclidean canonical gravity}
\label{sec:2.1}

3-dimensional Euclidean canonical gravity in the Ashtekar
formalism is defined over an oriented 2-manifold $\cal S$ which is
a foliation of the 3-manifold $\Si={\cal S}\times {\rm R}$
\cite{AL95,Th97a}. The basic variables are $SU(2)$ connections
$A_a^i$ and conjugate electric fields $E^b_j:=\e^{ba}e_{aj}$,
where we use $a,b,\dots =1,2$ for spatial indices and $i,j,\dots
=1,2,3$ for internal $SU(2)$ indices, $\e^{ab}$ is the
2-dimensional Levi-Civita tensor density of weight 1, and $e_a^j$
denotes the pull-back of the co-triads ${}^{(3)}e_a^i$ on $\cal
S$. The 2-metric on $\cal S$ reads $q_{ab}=e_a^i e_{bi}$. The
symplectic structure is given by $ \{A_a^i(x),
E^b_j(y)\}=\d^i_j\d^b_a\d^2(x,y)$, where we set the usual
gravitational constant $k_G=1$. Besides the Gauss and
diffeomorphism constraints, the dynamics is reflected by the
Ashtekar's Hamiltonian constraint \be \label{1} H(N):={1\over
2}\int_{\cal S}d^2x N\e^{ijk}F_{abi}{E^a_j E^b_k \over \sqrt{q}},
\ee where $F_{abi}$ is the curvature of $A_a^i$, $q$ denotes the
determinant of $q_{ab}$, and the smear function $N$ is some scalar
field.

Given any graph $\G$ with $n$ edges $e_I$, $I=1,\cdots,n$, and $m$
vertices $v_\b$, $\b=1,\cdots,m$, embedded in the 2-manifold $\cal
S$, the holonomy of the $SU(2)$ connection $A_a^i$ along any edge
$e_I$ gives an element of $SU(2)$ as: $h[A,e_I]={\cal
P}exp\int_{e_I}ds\dot{e}_I(s)A_a^i(e_I(s))\t_i$, where ${\cal P}$
denotes path ordering and $\t_i$ are the $SU(2)$ generators in the
fundamental representation. Given a function $f_n:
[SU(2)]^n\rightarrow \C$, the cylindrical function with respect to
$\G$ is defined as: $\Psi_{\G_n,f_n}(A):=f_n(h[e_1], \dots,
h[e_n])$. Since any two cylindrical functions based on different
graphs can always be viewed as being defined on the same graph
which is just constructed as the union of the original ones, it is
straightforward to define a scalar product for them by:
\be\label{b3} <\Psi_{\G_n,f_n}|\Psi_{\G_n,g_n}>:= \int_{[SU(2)]^n}
dh_1\dots dh_n f_n^{*}(h_1,\dots, h_n)g_n(h_1, \dots, h_n),\ee
where $dh_1\dots dh_n$ is the Haar measure of $[SU(2)]^n$ which is
naturally induced by that of $SU(2)$. The kinematical Hilbert
space, ${\cal H}_E=L_2(\overline{{\cal A}},d\mu_H)$, is obtained
by completing the space of all finite linear combinations of
cylindrical functions in the norm induced by the quadratic form
(\ref{b3}) on a cylindrical function.

The formal expression of the momentum operator is some functional
derivative with respect to $A_a^i$, i.e., $\hat{E}^a_i(x)=-i\hbar
(\d/ \d A_a^i(x))$. Its action on a cylindrical function yields
\be\label{b5} \hat{E}^a_i(x)\cc
\Psi_{\G_n,f_n}=-il_p\sum_{I=1}^n\int_{e_I}ds\dot{e}_I^a(s)
\d^2(x,e_I(s))X^I_i(t)\cc\Psi,\ee where $l_p=k_G\hbar$ is the
Planck length and $X^I_i(t)\equiv
Tr\left(h_I[0,t]\t_ih_I[t,1](\partial/\partial h_I)\right)$.
Classically, let \be \label{4.1} E^i\equiv {1\over 2}
\e^{ijk}\e_{ab}E^a_j E^b_k,\ee the area of any bounded region
$B\subset \cal S$ reads\be \label{4.2} V_B=\int_B d^2x
\sqrt{q}=\int_B d^2x \sqrt{E^iE_i}.\ee Note that $\hat{E}^a_i(x)$
is an operator-valued distribution rather than a genuine operator.
In order to construct a well-defined operator $\hat{V}_B$
corresponding to the classical area $V_B$, some regularization
procedure is necessary. The final expression of the area operator
reads \cite{Th97a} \be \label{b6} \hat{V}_B=\sum_{v_{\b}\in
B}\hat{V}_{v_{\b}}, \ee where \be \label{b7} \hat{V}_{v_\a}={l_p^2
\over
8}\sqrt{\sum_{I_{\a},J_{\a}}2sgn(e_I,e_J)X^I_{[i}X^J_{j]}X_I^iX_J^j},
\ee here $X^I_i\equiv X^I_i(0)=Tr\left(\t_ih_I(\partial/\partial
h_I)\right)$ is the right invariant vector field on $SU(2)$
evaluated at $h_I$ and $sgn(e_I,e_J)$ is the sign of
$\e_{ab}\dot{e}_I^a(0) \dot{e}_J^b(0)$. Note that, for
convenience, each edge is subdivided into two parts and equipped
each part with an orientation that is outgoing from the vertex.

The regularization of the Hamiltonian constraint operator is
rather complicated. By choosing the triangulation $T$ adapted to
the graph $\G$, a densely defined operator corresponding to $H(N)$
could be constructed as: \be \label{b8} \hat{H}(N)={2\over
\hbar}\sum_{v_{\b}\in \G}\sum_{\triangle,\triangle'\in T}
\e^{ij}\e^{kl}N(v)Tr\left(h_{\a_{ij}(\triangle')}h_{s_k(\triangle)}
[h_{s_k(\triangle)}^{-1},\sqrt{\hat{V}_v}]h_{s_l(\triangle)}
[h_{s_l(\triangle)}^{-1},\sqrt{\hat{V}_v}]\right).\ee We refer to
\cite{Th97a} for details.

A complete orthonormal basis in ${\cal H}_E$ is the
(non-gauge-invariant) spin network states. Moreover, the gauge
invariant spin network states, $\Psi_S(A)$, form a complete
orthonormal basis in the $SU(2)$ gauge invariant Hilbert space
${\cal H}_E^0=L_2(\overline{{\cal A}/{\cal G}},d\mu_H)$
\cite{RS95a,Ba96,AL95}. Let $\Phi$ denote the space of finite
linear combinations of functions $\Psi_S(A)$, the distributional
dual, $\Phi'$, is the set of all continuous linear functionals on
$\Phi$. We thus have the inclusion $\Phi\subset {\cal
H}_E^0\subset \Phi'$. A diffeomorphism invariant distribution on
$\Phi$ is defined by \be \label{b9} [\Psi_S]:=\sum_{\Psi\in
\{\Psi_S\}}\Psi,\ee where $\{\Psi_S\}\equiv\{\hat{U}(\varphi)\cc
\Psi_S, \varphi\in Diff({\cal S})\}$ is the orbit of $\Psi_S$
under the diffeomorphism group on $\cal S$. Therefore every
diffeomorphism invariant state is a linear combination of the
distributions $[\Psi_S]$ and belongs to $\Phi'$. This space of the
solutions to the diffeomorphism constraint is denoted by
$\Phi'_{Diff}$ \cite{AL95,Th97a}.

\subsection{Higgs field coupled to gravity}
\label{sec:2.2}

Suppose Higgs scalar fields $\phi(x)$ are valued in the Lie
algebra, $Lie(G)$, of some compact group $G$. The quantization of
$\phi=\phi_A \tau^A$, where $\tau^A$ are the generators of $G$,
could be included into the formalism of loop quantum gravity by
introducing the concept of point holonomies \cite{Th97b}.
Analogously to the holonomy associated to an edge $e_I$, a point
holonomy associated with a point $v$ is a $G$-valued function on
the space, $\cal U$, of Higgs fields given by \be \label{b10}
\underline{h}_v(\phi):=exp[\phi_A(v)\tau^A]. \ee A function
$\Psi(\phi)$ is said to be cylindrical with respect to a vertex
set $\{v_\beta\}$ if it depends only on the finite number of point
holonomies, i.e., \be \label{b11}
\Psi_{v,f}:=f[\underline{h}_{v_1}(\phi),\cdots,\underline{h}_{v_m}(\phi)],
\ee where f is a complex-valued function on $G^m$.

In quite analogy with Eq.(\ref{b3}), an inner product could be
defined on the space of cylindrical functions $\Psi_{v,f}$ with
the aid of Haar measure, and the Hilbert space, ${\cal
H}_S=L_2(\overline{\cal U},d\mu_H)$, of Higgs fields is obtained
by completing the space of all finite linear combinations of
cylindrical functions in the norm induced by this inner product.
Moreover, one could mimic the concept of spin networks to define
vertex functions on ${\cal U}$ by assigning each vertex $v_\beta$
an irreducible representations of $G$. It turns out that the
vertex functions provide an orthonormal basis for ${\cal H}_S$.
See \cite{Th97b} for details. The formal expression of the
momentum operator corresponding to the conjugate momentum, $p^A$,
of the Higgs fields $\phi_A$ reads
$\hat{p}^A(x)=-i\hbar\delta/\delta \phi_A(x)$. In the case of 2+1
space-times introduced in the last subsection, we can smear it on
two-surface ${\cal S}$ and get a densely well-defined operator
$\hat{p}^A({\cal S})\equiv \int_{{\cal S}}d^2x\hat{p}^A(x)$. Its
action on a cylindrical function yields as same as the result in
3+1 dimension \cite{Th97b}, \be \label{b12} \hat{p}^A({\cal
S})\circ\Psi_{v,f}=-i\hbar\sum_{v_\b}X_{v_\b}^A\circ\Psi_{v,f},
\ee where $X_{v_\b}^A\equiv
(1/2)[X_R^A(\underline{h}_{v_\b})+X_L^A(\underline{h}_{v_\b})]$,
here $X_R^A(\underline{h})$ and $X_L^A(\underline{h})$ are
respectively the right and left invariant vector fields on $G$.

In the framework of Higgs scalar fields coupled to gravity, the
elementary excitations of Higgs fields are at the vertexes of a
given graph, while those of gravity are along the edges. The total
Hilbert space is given by the product ${\cal H}_E\otimes{\cal
H}_S$. Moreover, in the case of 3+1 Lorentz gravity coupled to
Higgs scalar fields, through suitable regularization a densely
well-defined operator corresponding to the Hamiltonian of the
system could be constructed, which means the Hamiltonian is
completely finite without renormalization \cite{Th97b}.

\section{Dimension reduction}
\label{sec:3}

It is well known that 3+1 gravity with a hypersurface-orthogonal
killing vector field is equivalent to 2+1 gravity coupled to a
massless scalar field, since the early work of Geroch \cite{Ge71}.
While that work concerned only the equations of motion,
We now start with the variation principle to study static spacetimes.
Consider the Einstein-Hilbert action for 4-dimensional
Lorentz gravity, $S_{EH}[g^{ab}]=\int_M{\cal L}_G=\int\sqrt{-g}{
}^{(4)}R$, defined on a 4-manifold $M=\Sigma\times{\rm R}$. The
Lagrangian density can also be written as a version depending only
on the geometrical quantities on the hypersurface $\Sigma$ as \be
\label{c1} {\cal L}_G=\sqrt{h}{\cal N}(R+K_{ab}K^{ab}-K^2), \ee
where $h$ denotes the determinant of the induced 3-metric $h_{ab}$
on $\Sigma$, $R$ and $K_{ab}$ are respectively the scalar
curvature of $h_{ab}$ and extrinsic curvature of $\Sigma$,
$K\equiv K^a_a$, and $\cal N$ is the lapse function. If we only
consider the configurations of static space-times, it is
convenient to choose the hypersurface $\Sigma$ orthogonal to the
timelike killing vector field $\xi^a$. Then the extrinsic
curvature of $\Sigma$ vanishes, and Eq.(\ref{c1}) is reduced to
${\cal L}_G={\cal N}\sqrt{h}R$.

We now conformally transform $h_{ab}$ as $\bar{h}_{ab}=\Omega^{-2}
h_{ab}$, and hence obtain \cite{Wa84} \be \label{c2}
R=\Omega^{-2}[\bar{R}-4\bar{h}^{ab}\bar{\nabla}_a\bar{\nabla}_b\ln\Omega
-2\bar{h}^{ab}(\bar{\nabla}_a\ln\Omega)\bar{\nabla}_b\ln\Omega],
\ee where $\bar{R}$ is the scalar curvature of $\bar{h}_{ab}$.
Taking account of $\bar{h}=\Omega^{-6}h$, we let $\Omega={\cal
N}^{-1}$. The Lagrangian density then becomes \be \label{c3} {\cal
L}_G=\sqrt{\overline{h}}[\bar{R}-4\bar{h}^{ab}
\bar{\nabla}_a\bar{\nabla}_b\ln\Omega
-2\bar{h}^{ab}(\bar{\nabla}_a\ln\Omega)\bar{\nabla}_b\ln\Omega].
\ee Note that the second term in Eq.(\ref{c3}) is a total
divergence term, since $\bar{\nabla}_a$ is compatible with
$\bar{h}^{bc}$. Let $\Lambda\equiv \sqrt{2}\ln\Omega$,
straightforward calculations show that the Lagrangian (\ref{c3})
gives the same equations of motion as those of Euclidean 3-gravity
$\bar{h}^{ab}$ coupled to a massless Klein-Gordon field $\Lambda$,
which is defined by the coupled action \be \label{c4}
S_E+S_{KG}=\int_{\Sigma}\sqrt{\bar{h}}
[\bar{R}-\bar{h}^{ab}(\partial_a\Lambda)\partial_b\Lambda]. \ee
Therefore, a static 4-dimensional space-time is "conformally"
equivalent to 3-dimensional Euclidean gravity coupled to a
massless scalar field.

The above dimensional reduction motivates us to quantize the model
of 3-dimensional Euclidean gravity coupled to massless
Klein-Gordon fields as an equivalent description of quantized
static space-times. In order to apply the canonical quantization
framework of loop quantum gravity, we "imaginarize" the scalar
field as $\phi=i\Lambda$, and write the gravitational action in
Eq.(\ref{c4}) in Palatini formalism. The total action is then
defined as \be \label{c5} S_T({ }^{(3)}\bar{e},{
}^{(3)}A,\phi)=S_P+{1\over 2}S_{KG}={1\over
2}\int_{\Sigma}[\epsilon^{abc}{ }^{(3)}\bar{e}_{ai}{
}^{(3)}F_{bc}^i
+\sqrt{\bar{h}}\bar{h}^{ab}(\partial_a\phi)\partial_b\phi], \ee
where ${ }^{(3)}F_{bc}^i$ is the curvature of the $SU(2)$
connection 1-form, ${}^{(3)}A_a^i$, on $\Sigma$. In complete
analogy with the Palatiti formalism coupled to real Klein-Gordon
fields \cite{Ro93}, the variation of action (\ref{c5}) gives the
same equations of motion as those of action (\ref{c4}). Hence, the
two actions give the same classical theory. Suppose the 3-manifold
admit a foliation $\Sigma={\cal S}\times{\rm R}$. In the
corresponding Hamiltonian formalism the above imaginarization is
just a canonical transformation on the phase space. Through 2+1
decomposition one can obtain the Hamiltonian of the system, which
is just the linear combination of following first class
constraints: \be\label{5.1} G(\Lambda^i):=\int_{\cal S}d^2x
\Lambda^i{\cal D}_a \bar{E}^a_i, \ee \be\label{5.2}
V(N^a):=-\int_{\cal S}N^a(\bar{E}^b_iF_{ab}^i+p\partial_a\phi),
\ee \be \label{c6} H(N):={1\over 2}\int_{\cal S}d^2x {N\over
\sqrt{\bar{q}}}[\e^{ijk}F_{abi}\bar{E}^a_j \bar{E}^b_k -
\bar{E}^a_i \bar{E}^{bi}(\partial_a\phi)\partial_b\phi - p^2], \ee
where $p$ is the conjugate momentum of $\phi$. They are the Gauss,
vector, and Hamiltonian constraints. The former two constraints
reflect the symmetries of internal $SU(2)$ and the diffeomorphisms
on ${\cal S}$. The latter one reflects dynamics. Note that the
2-metric $\bar{q}_{ab}$ induced from $\bar{h}_{ab}$ is related to
that from $h_{ab}$ by\be \label{6.1}
\bar{q}_{ab}=\Omega^{-2}q_{ab}.\ee

\section{Canonical quantization}
\label{sec:4}

As purely imaginary numbers, the scalar fields $\phi$ defined in
last section are valued in the Lie algebra of $U(1)$. We are now
ready to apply the canonical quantization framework outlined in
last section to quantize the model. Although here $\phi$
originally are not Higgs fields, we may still suppose them to be
located at the vertexes of a graph, as the same treatment appears
in the study of quantum fields on a lattice \cite{MM97}. Thus, the
method to quantize Higgs fields could be naturally borrowed. The
kinematical Hilbert space is still given by the product ${\cal
H}_E\otimes{\cal H}_S$, while the cylindrical functions in ${\cal
H}_S$ are defined on $U(1)^m$ which is the product of point
holonomies associated to the vertexes $v_\b$. The Gauss and
diffeomorphism constraints could be solved by exactly the same
procedure employed for pure gravity. The non-trivial task would be
how to construct a well-defined operator corresponding to
Eq.(\ref{c6}). It turns out that by introducing properly the
triangulation $T$ of ${\cal S}$ adapting to the given graph
$\Gamma$, for example according to Ref.\cite{Th97a}, the
Hamiltonian constraint can be regulated in a consistent strategy
and promoted to a densely defined operator. While the
gravitational term has been expressed as Eq.(\ref{b8}), we now
regulate the other two terms involving the scalar fields. From
Eq.(\ref{4.1}), we have $\bar{q}=\bar{E}^i\bar{E}_i$ and
\cite{Th97a} \be \label{c7} \bar{E}_i ={1\over
2}\e^{ab}\e_{ijk}\{A_a^j,\bar{V}\}\{A_b^k,\bar{V}\},\ee where
$\bar{V}$ is the area of ${\cal S}$ measured by $\bar{q}_{ab}$.
Let $\bar{V}(x,\e)\equiv \int d^2y \theta_\e(x,y)\sqrt{\bar{q}}$,
where $\theta_\e(x,y)$ is the characteristic function of a box of
coordinate size $\e^2$ and center $x$. Consider the following
regulated point-splitting of the term in Eq.(\ref{c6}) which
involves the momentum $p$, \begin{eqnarray} \label{c8}
H_{KG,p}^{\e}(N)={1\over 2}\int d^2x N(x)p(x)\int d^2y
p(y)\theta_\e(x,y)\theta_\e(u,x)\theta_\e(w,y) \nonumber\\
\int d^2u\left[{{\bar{E}^i}\over
{\left(\bar{V}(u,\e)\right)^{3/2}}}\right] \int
d^2w\left[{{\bar{E}_i}\over
{\left(\bar{V}(w,\e)\right)^{3/2}}}\right]
\nonumber\\
= {1\over 4}\int d^2x N(x)p(x)\int d^2y p(y)\theta_\e(x,y)\int
d^2u\int d^2w\theta_\e(u,x)\theta_\e(w,y) \nonumber\\
\e^{ab}(u)\{A_a^j(u),\sqrt[4]{\bar{V}(u,\e)}\}
\{A_b^k(u),\sqrt[4]{\bar{V}(u,\e)}\} \nonumber\\
\e^{cd}(w)\{A_{cj}(w),\sqrt[4]{\bar{V}(w,\e)}\}
\{A_{dk}(w),\sqrt[4]{\bar{V}(w,\e)}\}
\nonumber\\
=\int d^2x N(x)p(x)\int d^2y p(y)\theta_\e(x,y)\int d^2u\int
d^2w\theta_\e(u,x)\theta_\e(w,y)
\nonumber\\
\e^{ab}(u)Tr[\{A_a(u),\sqrt[4]{\bar{V}(u,\e)}\}
\{A_{c}(w),\sqrt[4]{\bar{V}(w,\e)}\}] \nonumber\\
\e^{cd}(w)Tr[\{A_b(u),\sqrt[4]{\bar{V}(u,\e)}\}
\{A_{d}(w),\sqrt[4]{\bar{V}(w,\e)}\}] , \end{eqnarray} where
$A_a\equiv A_a^i\t_i$ and the equation $Tr(\t_i\t_j)=-\d_{ij}/2$
is used. Let $\triangle$ be a triangle of the triangulation $T$
adapted to the graph $\G$ and its basepoint be a vertex
$v(\triangle)$ of $\G$. As the image of $[0,\d]$, where $\d$ is a
small parameter, the two edges $s_I(\triangle), I=1,2$ incident at
$v(\triangle)$ coincide with the segments of two edges of $\G$. In
the light of the observation in Ref.\cite{Th97b}, we have \fa
\label{c9} \theta_\e(x,y)\e^{IJ}h_{s_I(\triangle)}
\{h_{s_I(\triangle)}^{-1},\sqrt[4]{\bar{V}(v(\triangle),\e)}\}
h_{s_J(\triangle)}
\{h_{s_J(\triangle)}^{-1},\sqrt[4]{\bar{V}(v(\triangle),\e)}\} \nonumber\\
=2\int_{\triangle}d^2y \theta_\e(x,y)
\e^{ab}(y)\{A_a(y),\sqrt[4]{\bar{V}(y,\e)}\}
\{A_b(y),\sqrt[4]{\bar{V}(y,\e)}\}+O(\d^3). \ffa Thus, up to order
$\d$ which vanishes in the limit as we remove the triangulation,
Eq.(\ref{c8}) can be expressed as \fa \label{c10}
H_{KG,p}^{T,\e}(N)=\int d^2x N(x)p(x)\int d^2y
p(y)\theta_\e(x,y)\sum_{\triangle,\triangle'\in
T}{1\over 4}\theta_\e(v(\triangle),x)\theta_\e(v(\triangle'),y) \nonumber\\
\e^{IJ}Tr\left(h_{s_I(\triangle)}
\{h_{s_I(\triangle)}^{-1},\sqrt[4]{\bar{V}(v(\triangle),\e)}\}
h_{s_K(\triangle')}
\{h_{s_K(\triangle')}^{-1},\sqrt[4]{\bar{V}(v(\triangle'),\e)}\}\right) \nonumber\\
\e^{KL}Tr\left(h_{s_J(\triangle)}
\{h_{s_J(\triangle)}^{-1},\sqrt[4]{\bar{V}(v(\triangle),\e)}\}
h_{s_L(\triangle')}
\{h_{s_L(\triangle')}^{-1},\sqrt[4]{\bar{V}(v(\triangle'),\e)}\}\right).
\ffa Eq.(\ref{b12}) implies $\hat{p}({\cal
S})=-i\hbar\sum_{v_{\b}}X_{v_{\b}}$, where $X_v\equiv
1/2[X_R(\underline{h}_v)+X_L(\underline{h}_v)]$, here
$X_R(\underline{h}_v)$ and $X_L(\underline{h}_v)$ are respectively
the right and left invariant vector fields at $\underline{h}_v\in
U(1)$. Now we replace $\int d^2x p(x)$ and $\bar{V}$ by their
corresponding operators, replace Poisson brackets by commutators
times $1/(i\hbar)$, and take $\e$ to zero. The result reads \fa
\label{b13} \hat{H}_{KG,p}^T(N)= -{1\over
4\hbar^2}\sum_{v_\b}N(v_\b)X_{v_{\b}}X_{v_{\b}}
\sum_{\triangle(v_{\b}),\triangle'(v_{\b})\in T}\e^{IJ}\e^{KL} \nonumber\\
Tr\left(h_{s_I(\triangle)}
[h_{s_I(\triangle)}^{-1},\sqrt[4]{\hat{\bar{V}}(v)}]
h_{s_K(\triangle')}
[h_{s_K(\triangle')}^{-1},\sqrt[4]{\hat{\bar{V}}(v)}]\right) \nonumber\\
Tr\left(h_{s_J(\triangle)}
[h_{s_J(\triangle)}^{-1},\sqrt[4]{\hat{\bar{V}}(v)}]
h_{s_L(\triangle')}[h_{s_L(\triangle')}^{-1},\sqrt[4]{\hat{\bar{V}}(v)}]\right),\ffa
which is a densely well-defined operator. Note that
$\hat{\bar{V}}(v)$ is expressed as the same as Eq.(\ref{b7}). Now
we turn to the term, $H_{KG,\phi}$, involving the derivatives of
$\phi$ and regulate it as \fa \label{c14} H_{KG,\phi}^\e(N) =
{1\over 2}\int d^2x N(x)\int d^2y
\theta_\e(x,y)\e^{ab}(x)(\partial_a\phi(x)){e_{bi}(x)\over
\sqrt{\bar{V}(x,\e)}}\e^{cd}(y)(\partial_c\phi(y)){e_{d}^i(x)\over
\sqrt{\bar{V}(y,\e)}} \nonumber\\
=-\int d^2x N(x)\int d^2y
\theta_\e(x,y)\e^{ab}(x)(\partial_a\phi(x))\e^{cd}(y)(\partial_c\phi(y))
\nonumber\\
Tr\left(\{A_b(x),\sqrt{\bar{V}(x,\e)}\}\{A_d(y),\sqrt{\bar{V}(y,\e)}\}\right).
\ffa Notice that classically we have, on an edge $s_I$ incident at
a vertex $v=s(0)$, \fa \label{c15}
\underline{h}^{-1}(v)[\underline{h}(s(\d))-\underline{h}(v)] &=&
\underline{h}^{-1}(v)\left[\exp\left(\phi(v)+\d\dot{s}^a(0)\partial_a\phi(v)+O(\d^2)\right)
-\underline{h}(v)\right] \nonumber\\
&=& \d\dot{s}^a(0)\partial_a\phi(v). \ffa Hence on the
triangulation $T$, Eq.(\ref{c14}) can be expressed as \fa
\label{c16} H_{KG,\phi}^{T,\e}(N)=-{1\over 4}
\sum_{\triangle,\triangle'\in T}N(v(\triangle))\e^{IJ}
\underline{h}^{-1}(v(\triangle))
\left(\underline{h}(s_I(v(\triangle)))-\underline{h}(v(\triangle))\right) \nonumber\\
\e^{KL}\underline{h}^{-1}(v(\triangle'))
\left(\underline{h}(s_K(v(\triangle')))-\underline{h}(v(\triangle'))\right)
\theta_\e(v(\triangle),v(\triangle')) \nonumber\\
Tr\left(h_{s_J(\triangle)}
\{h_{s_J(\triangle)}^{-1},\sqrt{\bar{V}(v(\triangle),\e)}\}
h_{s_L(\triangle')}
\{h_{s_L(\triangle')}^{-1},\sqrt{\bar{V}(v(\triangle'),\e)}\}\right).
\ffa In the limit $\e\rightarrow 0$, the operator version of Eq.
(\ref{c16}) reads \fa \label{c18} \hat{H}_{KG,\phi}^T(N)= {1\over
4\hbar^2}\sum_{v_\b}N(v)\underline{h}^{-2}(v)
\sum_{\triangle(v_{\b}),\triangle'(v_{\b})\in
T}\left(\underline{h}(s_I(v(\triangle)))-\underline{h}(v)\right)
\left(\underline{h}(s_K(v(\triangle')))-\underline{h}(v)\right) \nonumber\\
\e^{IJ}\e^{KL}Tr\left(h_{s_J(\triangle)}
[h_{s_J(\triangle)}^{-1},\sqrt{\hat{\bar{V}}(v)}]
h_{s_L(\triangle')}
[h_{s_L(\triangle')}^{-1},\sqrt{\hat{\bar{V}}(v)}]\right),\ffa
which is also densely well defined. In conclusion, the Hamiltonian
constraint (\ref{c6}) has been quantized as a densely defined
operator in ${\cal H}_E\otimes{\cal H}_S$.

\section{Future directions: area operator and beyond}
\label{sec:5}

The consistency of the Hamiltonian operator constructed in last
section is left for future investigations. While, from the
structure of this operator, it is reasonable to expect that it
should share the advantages of its analogue in 3+1 dimension
\cite{Th97b}, namely cylindrical consistency, diffeomorphism
covariance, and anomaly-freeness. Also, the complete set of
solutions to the all constraints can be characterized following
the procedure of Ref.\cite{Th97b}.

We now discuss the construction of an area operator. It should be
noted that the sum of the area operator associated to a vertex
$\hat{\bar{V}_v}$ used in last section does not correspond to the
area measured by the physical 2-metric $q_{ab}$. Classically, from
Eqs.(\ref{4.2}) and (\ref{6.1}) the physical area is\be \label{d1}
V_B=\int_B d^2x\Omega^2\sqrt{\hat{q}}=\int_B
d^2x\exp\left(-i\sqrt{2}\phi\right)\sqrt{\hat{q}}.\ee Let
$\chi\equiv i\sqrt{2}\phi$ and assume $\Omega\leq 1$ (the opposite
case is yet to be studied), we then have the following Fourier
transform \fa \label{d2} e^{-\chi}&=&{1\over \pi}\int_0^\infty
d\eta \left({1\over
1+\eta^2}\right)\left(e^{-i\eta\chi}+e^{i\eta\chi}\right)\nonumber\\
&=&{1\over \pi}\int_0^\infty d\eta \left({1\over
1+\eta^2}\right)\left(e^{\sqrt{2}\eta\phi}+e^{-\sqrt{2}\eta\phi}\right)\nonumber\\
&\equiv&{1\over \pi}\int_0^\infty d\eta \left({1\over
1+\eta^2}\right)(\underline{h}_x(\eta)+\underline{h}_x^{-1}(\eta)),
\ffa where $\underline{h}_x(\eta)\in U(1)$ depends on the
parameter $\eta$. Thus, taking account of Eq.(\ref{b6}) and
transform (\ref{d2}) it is straightforward to define an operator
corresponding to the area $V_B$ as\be \label{d3}
\hat{V}_B=\sum_{v_\b\in B}{1\over \pi}\int_0^\infty d\eta
\left({1\over
1+\eta^2}\right)(\underline{h}_{v_\b}(\eta)+\underline{h}_{v_\b}^{-1}(\eta))
\hat{\bar{V}}(v_\b).\ee This operator is not only densely defined
but also essentially self-adjoint and positive semi-definite. To
see the latter, notice that $\underline{h}_v(\eta)$ is a family of
unitary matrix and the adjointness relation implemented in ${\cal
H}_S$ is $\hat{\underline{h}}_v^\dag=\hat{\underline{h}}_v^{-1}$,
and hence \be \label{3.1}
\left(\hat{\underline{h}}_v(\eta)+\hat{\underline{h}}_v^{-1}(\eta)\right)^\dag
=\left(\hat{\underline{h}}_v^{-1}(\eta)+\hat{\underline{h}}_v(\eta)\right).\ee
With a well-defined area operator at hand, it is possible to
address some physical problems.

As argued in Ref.\cite{HH95}, the surface term arising from the
gravitational action of (\ref{c1}) could be taken as the
definition of the total energy even for space-times that are not
asymptotically flat. The derivation of the surface term depends
crucially on a reference background space-time, which is supposed
to be static. It is essentially expressed as \cite{HH95} \be
\label{d4} E=-\int_{{\cal S}_t^\infty}\sqrt{q}{\cal N}({
}^{(2)}K-{ }^{(2)}K_0),\ee where ${}^{(2)}K$ and ${}^{(2)}K_0$ are
the traces of the 2-dimensional extrinsic curvature of ${\cal
S}_t^\infty$ in $\Sigma_t$ corresponding respectively to the field
metric and the background metric; here the 2-surface ${\cal
S}_t^\infty$ is the intersection of $\Sigma_t$ and a boundary near
infinity. If we only consider static space-times, it seems
naturally to apply our framework of quantization and consider the
Hilbert space ${\cal H}_E\otimes{\cal H}_S$ defined on ${\cal
S}_t^\infty$. A proper regularization of Eq.(\ref{d4}) is
necessary before its quantization, and the area operator
(\ref{d3}) is supposed to play a key role \cite{Ma}. Another
appealing topic which deserves investigating is to calculate black
hole entropy. Our framework provides a possibility to count the
numbers of quantum states in the physical Hilbert space associated
to the apparent horizons of static black holes. Two essential
factors are needed for this consideration. First, we need a local
definition of apparent horizons, i.e., to define the horizon by
the intrinsic geometry of the 2-surface itself. This is in quite
analogy with the definition of isolated horizon which is a
generalization of event horizon \cite{AAK99}. Second, to solve
exactly all the quantum constaints. It would be amazing, if one
could find that the number of physical states becomes finite on an
apparent horizon while it is infinite on other non-horizon
surfaces.

\subsection*{Acknowledgements}
The author would like to acknowledge CGPG for hospitality at Penn
State and thank Abhay Ashtekar, Martin Bojowald, Steve Fairhurst,
Bruno Hartmann, Yi Ling for valuable and helpful discussions, and
the referee for helpful comments. This work is supported in part
by NSF grant PHY00-90091 and Eberly research funds of Penn State.

\end{document}